\documentclass[%
pra%
,reprint%
,twocolumn%
,secnumarabic%
,floatfix%
,superscriptaddress
,showpacs%
,amssymb, amsmath, nobibnotes, aps]{revtex4-1}

\usepackage[latin1]{inputenc}
\usepackage{epsfig}
\usepackage{color}

\usepackage[bookmarks=false]{hyperref}

\newcommand{\ind}[1]{_{\text{#1}}}
\newcommand{\bra}[1]{\left<#1\right|}
\newcommand{\ket}[1]{\left|#1\right>}

\newcommand{\tr}{\text{tr}}
\newcommand{\su}{\uparrow}
\newcommand{\sd}{\downarrow}

\begin{document}

\title{Relating Out-of-Time-Order Correlations to Entanglement via Multiple-Quantum Coherences}

\author{Martin G\"arttner}
\affiliation{JILA, NIST and the University of Colorado, Boulder, Colorado 80309, USA}
\affiliation{Kirchhoff-Institut f\"ur Physik, Universit\"at Heidelberg, Im Neuenheimer Feld 227, 69120 Heidelberg, Germany}
\author{Philipp Hauke}
\affiliation{Institute for Theoretical Physics, University of Innsbruck, 6020 Innsbruck, Austria}
\affiliation{Institute for Quantum Optics and Quantum Information of the Austrian Academy of Sciences, 6020 Innsbruck, Austria}
\affiliation{Kirchhoff-Institut f\"ur Physik, Universit\"at Heidelberg, Im Neuenheimer Feld 227, 69120 Heidelberg, Germany}
\author{Ana Maria Rey}
\affiliation{JILA, NIST and the University of Colorado, Boulder, Colorado 80309, USA}

\date{\today}

\begin{abstract}
Out-of-time-order correlations (OTOCs) characterize the scrambling, or delocalization, of quantum information over all the degrees of freedom of a system and thus have been proposed as a proxy for chaos in quantum systems. Recent experimental progress in measuring OTOCs calls for a more thorough understanding of how these quantities characterize
complex quantum systems, most importantly in terms of the buildup of entanglement.
Although a connection between OTOCs and entanglement entropy has been derived, the latter only quantifies entanglement in pure systems and is hard to access experimentally.
In this work, we formally demonstrate that the multiple-quantum coherence spectra, a specific family of OTOCs well known in NMR, can be used as an entanglement witness and as a direct probe of multiparticle entanglement. 
Our results open a path to experimentally testing the fascinating idea that entanglement is the underlying glue that links thermodynamics, statistical mechanics, and quantum gravity.
\end{abstract}


\maketitle

%
Entanglement in quantum systems is a resource for quantum computation and communication and has been called \emph{the} characteristic trait of quantum mechanics \cite{schroedinger1935}. Recently, it has also been proposed \cite{dAlessio2016} and experimentally tested in proof-of-principle experiments \cite{Kaufman2016,Neill2016} that quantum entanglement is in fact the key concept behind thermalization in isolated quantum systems. Essentially, the approach to equilibrium can be understood as the spreading of entanglement through the system's degrees of freedom. 
In parallel, the concept of ``scrambling'' in many-body systems, which refers to the delocalization of quantum information over all of a system's degrees of freedom, has gained great attention \cite{Hosur2016, Yao2016, Swingle2016, Rozenbaum2017, Zhu2016, Shen2016, Li2016, bohrdt2017, swingle2017}, motivated by the finding that special models with thermal states ``holographically dual'' to black holes can thermalize and scramble quantum information at the fastest rate allowed by nature \cite{Maldacena2015, Sekino2008}. 
The scrambling rate can be quantified through out-of-time-ordered correlators (OTOCs), which have been connected to entanglement via the R\'enyi entropy \cite{Hosur2016,Fan2017}. 
However, the R\'enyi entropy is a strict entanglement monotone only for pure systems and hard to access experimentally, requiring resources that scale exponentially with the subsystem size as well as single-particle addressing.
Therefore, it is desirable to establish experimentally accessible entanglement witnesses applicable to open as well as isolated quantum systems which can be used to quantify scrambling.

In this Letter, we formally show that a specific family of OTOCs, first developed in NMR under the name of the multiple-quantum coherence (MQC) spectra, are useful entanglement witnesses. The MQC protocol has been known for many years to  be a suitable method to  quantify the development of many-body quantum coherences \cite{Baum1985, Cappellaro2014}. Recently, it has been applied to describe the spreading of correlations \cite{Baum1985, Baum1986, Munowitz1987, Sanchez2014} and as a signature of localization effects \cite{Alvarez2010, Alvarez2013, Alvarez2015, Wei2016}. While connections between MQCs and entanglement have been pointed out in Refs.~\cite{Doronin2003, Furman2008, Furman2009} and witnesses of two-particle entanglement have been constructed in Refs.~\cite{Feldman2008, Feldman2012}, to date a formal relation between the MQC spectrum and multiparticle entanglement generally applicable to mixed states does not exist. Here, we formally establish such a relation by deriving entanglement witnesses from the MQC intensities, as well as a relationship between MQCs and the quantum Fisher information (QFI) \cite{Toth2014, Pezze2016}, a well-known witness of multiparticle entanglement. 

To illustrate the power of these connections, we use the specific example of a long-range Ising model in a transverse field. We start the dynamics from a pure initial state, but show the applicability of the witness to mixed states by including decoherence arising from light scattering during the dynamics. This type of decoherence is relevant for a broad class of quantum systems. Our results demonstrate the existence of an experimentally accessible link between scrambling measured by OTOCs and entanglement, provided by the MQCs.

MQCs have a long tradition in NMR systems, which typically operate at high temperature. Measuring MQCs in pure and almost zero temperature initial states is now becoming feasible in cold-atom experiments, including Bose-Einstein condensates, ultracold atoms in cavities, or trapped ions \cite{Widera2008, Cucchietti2010, Leroux2010, Douglas2015, Linnemann2016, Swingle2016, Macri2016, Gaerttner2016}. Such experiments open the possibility to probe the rich information contained in an entangled state via MQCs.

%
We start by introducing the MQCs, which have been used as a means for quantifying quantum coherence \cite{Cappellaro2014, Baum1985, Munowitz1987}. Let $\ket{\psi_i}$ be the eigenstates of a Hermitian operator $\hat{A}$ and $\lambda_i$ the corresponding discrete eigenvalues. We divide the density matrix of an arbitrary state $\hat\rho$ into blocks as $\hat\rho=\sum_m \sum_{\lambda_i-\lambda_j=m}\rho_{ij}\ket{\psi_i}\bra{\psi_j}=\sum_m \hat\rho_m$. Thus, $\hat\rho_m$ contains all coherences between states with eigenvalues of $\hat A$ that differ by $m$. An experimentally accessible quantifier of these MQCs is the Frobenius norm $I_m(\hat\rho) =(\|\hat\rho_m\|_2)^2 =\tr[\hat\rho_{m}^\dagger \hat\rho_m]$ called multiple-quantum intensity. The key idea is that $I_m$ can be directly accessed in an experiment that has the ability to reverse the dynamics that created the state of interest $\hat\rho$ from an initial fiducial state $\hat\rho_0$. 
In this context, the  time reversal  can be connected with the  concept of  many-body Loschmidt echoes, well-known probes of irreversibility and chaos \cite{Rhim1970, Rhim1971, Pastawski2000, Zangara2015, Zangara2017}.

\begin{figure}[t]
  \centering
 \includegraphics[width=\columnwidth]{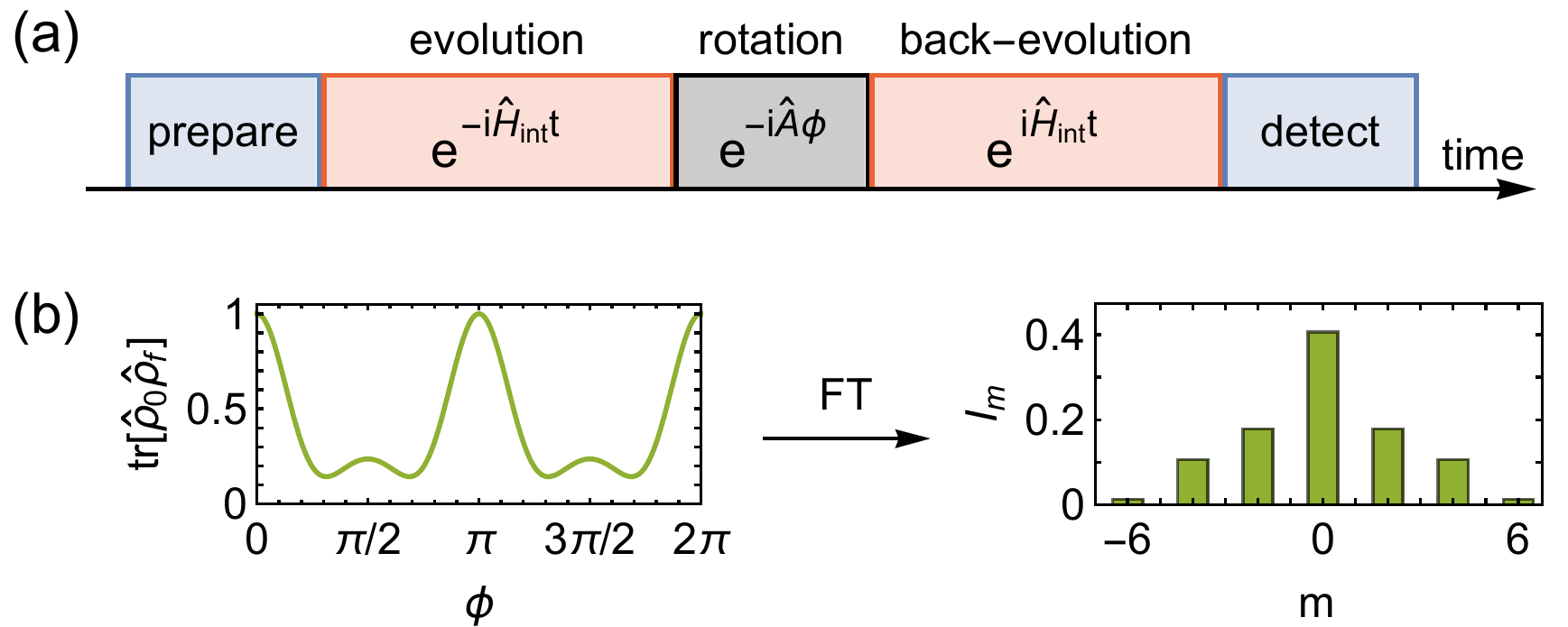}
\caption{(a) Illustration of the scheme for measuring the coherences using time reversal. The state of interest $\hat\rho_t$ is reached after the first evolution period. The rotation then imprints a phase $m\phi$ on each sector $\hat\rho_m$ of the density matrix (see text). Evolving backward and measuring the overlap with the initial state as a function of $\phi$, the coherences $I_m$ of $\hat \rho_t$ are retrieved as the Fourier components of this signal. (b) An example for the fidelity signal obtained from time evolution under the Ising Hamiltonian [Eq.~\eqref{eq:HTrIsing} with $\Omega=0$] and rotations about the $z$ axis of the spin ($\hat A=\hat S_z$). By Fourier transforming this signal, one obtains the intensities $I_m$, which quantify the magnitude of the $m$th order coherences of $\hat\rho_t$.}
\label{fig:scheme}
\end{figure}

The protocol to measure $I_m$ is as follows \cite{Baum1985, Gaerttner2016} (see Fig.~\ref{fig:scheme}): evolve $\hat\rho_0$ into $\hat \rho_t$ under a nontrivial unitary evolution $e^{-i\hat H\ind{int}t}$, apply $\hat W(\phi)=e^{-i\hat A\phi}$, evolve backward with $e^{i\hat H\ind{int}t}$ to $\hat\rho_f$, and finally measure the probability to find the system in the initial state $\tr[\hat\rho_0\hat\rho_f]$ (if $\hat\rho_0$ is pure, this is the fidelity).
Noting that $\hat W(\phi)\hat \rho_m \hat W^\dagger(\phi)=e^{im\phi}\hat\rho_m$ and using cyclic permutations under the trace, one finds
\begin{equation}
 F_t(\phi)\equiv \tr[\hat\rho_0\hat\rho_f] = \tr[\hat\rho_t \, \hat\rho_t(\phi)] = \sum_m I_m(\hat\rho_t) e^{-i m\phi},
  \label{eq:scheme}
 \end{equation}
where $\hat\rho_t(\phi) = \hat W(\phi) \hat \rho_t \hat W^\dagger(\phi)$.
Thus, by Fourier transforming the signal with respect to $\phi$, one obtains the MQC spectrum $\left\{I_m(\hat\rho_t)\right\}$ (see \cite{SM} for details). 
For NMR systems typically operating at infinite temperature, this overlap measurement reduces to a magnetization measurement, making it possible to observe coherences as high as $m \sim 7000$ \cite{Alvarez2013, Alvarez2015}. Nevertheless, the perturbative nature of the coherences present in highly mixed states, which facilitates experimental access of the MQCs, also implies that the underlying quantum complexity and entanglement content in those states are small in comparison to pure states. For pure states, measuring MQCs requires a fidelity measurement that encodes information about $N$-body correlations in an $N$-particle system \footnote{While in NMR the overlap measurement consists in measuring the magnetization $\sum_i\hat\sigma_i^z$, for pure states the projector $\bigotimes_i (1-\hat\sigma_i^z)$ needs to be measured, which involves spin correlations of any order.}.
Despite the fact that, in general, the resources required for measuring fidelity scale unfavorably with the system size, the feasibility of such a measurement has been demonstrated for up to 50 particles \cite{Gaerttner2016}, much beyond what is possible with schemes involving measuring entanglement entropy.

The connection between MQCs and OTOCs becomes apparent from $\hat\rho_t=e^{-i\hat H\ind{int}t}\hat\rho_0 e^{i\hat H\ind{int}t}$. By defining $\hat V_0=\hat\rho_0$, if $\hat V_0\hat\rho_0 = \hat\rho_0$ \footnote{This can be achieved easily for pure states and for strongly mixed states encountered in NMR experiments.}, the above expression can be recast as \cite{Gaerttner2016, Wei2016}
\begin{equation}
 F_t(\phi)\equiv \tr[\hat W_t^\dagger(\phi)\hat V^\dagger_0\hat W_t(\phi)\hat V_0 \rho_0]=\langle  W_t^\dagger(\phi)\hat V^\dagger_0\hat W_t(\phi)\hat V_0\rangle
  \label{eq:OTOC}
\end{equation}   
where  $\hat{W}_t(\phi)=e^{i \hat H\ind{int}t}\hat{W}(\phi) e^{-i \hat H\ind{int}t}$. $F_t(\phi)$ is therefore an OTOC function, a specific product of  Heisenberg operators not acting in normal order. When $\hat{W}(\phi)$ and $\hat{V}_0$ are chosen to be initially commuting operators, then $F_t(\phi)=1-\langle|[ \hat{W}_t(\phi), \hat{V}_0]|^2\rangle$. The growth of the norm of the commutator, i.e., the degree by which the initially commuting operators fail to commute at later times due to the many-body interactions generated by $\hat H\ind{int}$, is commonly used as an operational definition of the scrambling rate \cite{Hosur2016, Swingle2016, Yao2016}.  
Scrambling can be interpreted as the process by which the information encoded in the initial state, through the interactions, is distributed over the other degrees of freedom of the system. This process makes it no longer possible to retrieve the initial information by local operations and measurements.

%
We are now in the position to state the main results of the Letter.

First, the second moment of the MQC spectrum [$F_I/2$, defined in Eq.~\eqref{eq:boundQFI}] provides a lower bound on the quantum Fisher information $F_Q$
\begin{equation}
\begin{aligned}
F_I(\hat\rho_t,\hat A) & \equiv 2\sum_{m=-N}^N I_m(\hat\rho_t) m^2  \\
&=\left. -2\frac{\partial^2 F_t(\phi)}{\partial \phi^2}\right|_{\phi=0} \leq F_Q(\hat\rho_t,\hat A)
\end{aligned}
\label{eq:boundQFI}
\end{equation}
This expression becomes an equality for pure states $\hat\rho_t$.

The QFI has been introduced to quantify the maximal precision with which a parameter $\phi$ in the unitary $\hat{W}(\phi) = e^{-i\hat A\phi}$ can be estimated using the quantum state $\hat\rho$ as an input to an interferometer. It bounds the minimal variance of $\phi$ as $\Delta \phi \geq 1/\sqrt{F_Q(\hat\rho,\hat A)}$ (Cram\'er-Rao bound) \citep{braunstein1994}.
It has been shown that if $F_Q(\hat\rho,\hat A)> b_k \equiv n k^2 + (N-nk)^2$ (note that $b_k\geq Nk$), where $n$ is the integer part of $N/k$, then $\hat\rho$ is $(k+1)$-particle entangled \cite{Pezze2009, Hyllus2012, Toth2012}.
To derive expression \eqref{eq:boundQFI}, we used the relation  $ F_I(\hat\rho,\hat A)= 4\tr[\hat \rho^2 \hat A^2-(\hat\rho \hat A)^2]$, which is a lower bound on the QFI \cite{Yadin2016} (see also \cite{SM}). The choice of the generator $\hat A$ can be optimized for detecting the entanglement of a given state using intuition from quantum metrology.

The relation \eqref{eq:boundQFI}  has a number of implications, the most direct one being that $F_I$ inherits the property of $F_Q$ of being a witness for multiparticle entanglement; i.e., $F_I>b_k$ implies $F_Q>b_k$ and thus $(k+1)$-particle entanglement. This allows us to establish an intimate connection between scrambling of quantum information and the buildup of entanglement. Namely, the $\phi$ dependence of the OTOC $F_t(\phi)$ encodes information about the entanglement content of the state $\hat\rho_t$. We also note that, for thermal states, QFI can be directly related to dynamic susceptibilities, as demonstrated in  Refs.~\cite{Hauke2016} [see explicitly Eq.~(4)] and \cite{Pappalardi2017}, which are well-known signatures of quantum critical behavior and phase transitions. 
Moreover, the QFI is a measure of macroscopic coherences, such as appear in "cat states" \cite{Yadin2016}. 

Second, each individual $I_m$ by itself can be used as an  entanglement witness.
The quantity $F_I$ only characterizes the second moment of the MQC spectrum or, equivalently, only depends on the small-$\phi$ behavior of the measured observable $F_t(\phi)$, while the MQC spectrum, i.e., each individual $I_m$ contains much more detailed information about the state $\hat\rho$.
To show that  individual $I_m$ can witness entanglement, we use two properties \cite{SM}: First, the $I_m$ are convex, or nonincreasing under mixing [$I_m[p\hat\rho_1+(1-p)\hat\rho_2]\leq pI_m(\hat\rho_1)+(1-p)I_m(\hat\rho_2)$ for $m\neq 0$]. Second,  coherences of product states can be obtained from those of the constituent subensembles by $I_m(\hat\rho_A \otimes \hat\rho_B)=\sum_k  I_{m+k}(\hat\rho_A)I_{m-k}(\hat\rho_B)$. With these two properties, one can, for a given $m$, bound the maximal $I_m$ achievable on the set of separable states. 

In the following, we outline how to derive such bounds for systems of spin $1/2$ particles. The detailed proof can be found in the Supplemental Material \cite{SM}. 
The spins are described by Pauli operators $\hat{\sigma}_j^{\alpha}$, $\alpha=x,y,z$, $j=1,\dots, N$, with the eigenstates of $\hat{\sigma}_j^{z}$ denoted by $\ket{\su}_j$ and $\ket{\sd}_j$. 
We calculate the maximal $I_m$ achievable with a separable state.
Without loss of generality, we choose $\hat A=\hat S_z = \sum_j \hat{\sigma}_j^z/2$ \footnote{For a general local operator $\hat{A}=\sum_j \mathbf{n}_j\cdot \mathbf{\hat s}_j$, we can just define the basis of the Hilbert space as a product of the eigenstates of the local spin operators $\mathbf{n}_i\cdot \mathbf{\hat s}_j$, which formally maps the problem to the case we consider explicitly.}. 
It follows from the convexity that the maximum $I_m$ is assumed for pure states, which for separable states take the most general form $\bigotimes_j \left(\sqrt{p_j}\ket{\su}_j+e^{i\varphi_j} \sqrt{1-p_j}\ket{\sd}_j\right)$. 
From the rule for building tensor products, it follows that $I_m$ is independent of $\varphi_j$ and is a quadratic polynomial in the $p_j$. Noting that $I_m$ is invariant under $p_j\rightarrow 1-p_j$, the maximum is assumed when all $p_j$ are either extremal (zero or one) or equal to $1/2$. For such a state with $N_+$ spins in the equal superposition state ($p=1/2$), $I_m$ can be calculated analytically and optimized numerically with respect to $N_+$, which yields
\begin{equation}
\label{eq:bound}
 I_m^{\rm max, sep}= \max_{N_+\in \{0,\ldots, N\}} \frac{(2N_+)!}{4^{N_+}(N_+-m)!(N_++m)!} \, .
\end{equation}
Thus, if for a given state $\hat\rho$ and rotation generated by $\hat A$, one has $I_m>I_m^{\rm max, sep}$ for some $m$, then $\hat\rho$ must be entangled.
Note also that $I_N$ is a witness of genuine $N$-partite entanglement \cite{Leibfried2005}.

%
We now illustrate these results by applying them to the specific case of collective spin
models. We consider a system of $N$ spin $1/2$ particles and the coherences with respect to the collective spin operator $\hat A=\hat S_\mathbf{n}=\sum_j  \mathbf{\hat s}_j\cdot \mathbf{n}$, with $\mathbf{\hat s}_j=(\hat{\sigma}_j^x,\hat{\sigma}_j^y,\hat{\sigma}_j^z)/2$ and a unit vector $\mathbf{n}=(n^x,n^y,n^z)$. Thus, the spectrum of $\hat A$ consists of the (half) integers $M=-N/2,\ldots, N/2$, and we define the $m$th order coherence $\hat\rho_m$ as the block of the density matrix spanned by $\ket{\phi_M}\bra{\phi_{M+m}}$, where $\ket{\phi_M}$ are the eigenstates of $\hat A$ with eigenvalue $M$.
We study an all-to-all transverse-field Ising model
\begin{equation}
 \hat H\ind{int}=-J/N \hat S_x^2 - \Omega \hat S_z
 \label{eq:HTrIsing}
\end{equation}
where the spins are initially prepared in $\ket{\psi_0}=\ket{\uparrow}^{\otimes N}$. In the absence of decoherence, the dynamics is restricted to the symmetric Dicke manifold, which makes it very easy to numerically simulate the dynamics of large numbers of spins.

\begin{figure}[t]
  \centering
  \includegraphics[width=\columnwidth]{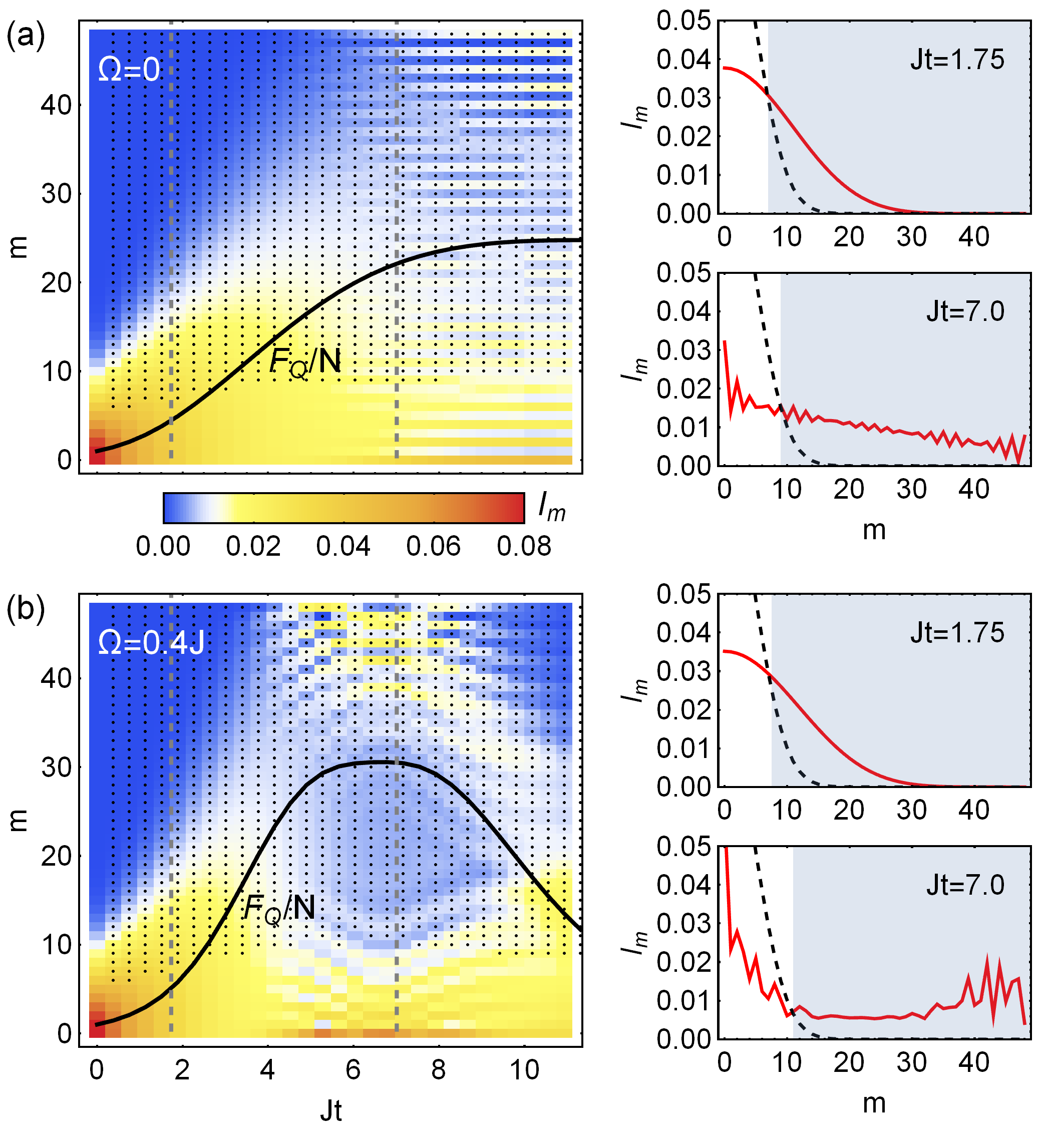}
\caption{
MQC spectra for evolution under the Ising (a) and transverse-field Ising (b) Hamiltonian as a function of the evolution time for $N=48$ spins. The QFI per particle is shown on top of the density plot as a solid line. $(k+1)$-particle entanglement is detected if $F_Q/N>k$ and in the pure case $F_Q=F_I$. The direction of the rotation axis $\mathbf{n}$ is optimized for each $t$. The pixels corresponding to those $I_m$ that violate the bound for separable states are marked with a dot. At late times reflection at the boundary of the MQC spectrum at $m=N$ leads to self-interference and fragmentation of the coherence spectrum. The right panels show the coherence spectrum (red solid) and entanglement bounds (black dashed) at specific times, indicated by the dashed lines in the left panels. The gray shading shows for which $m$ the bounds are violated.  
}
\label{fig:qfi_im_opt}
\end{figure}

In Fig.~\ref{fig:qfi_im_opt}, we illustrate the time evolution of the coherence spectrum $I_m$ for zero and nonzero transverse field. The QFI per particle, shown as a black line, is proportional to the variance of the coherence spectrum. The figure shows that the $I_m$ surpass the bounds for separable states in large parts of the spectrum. A complex pattern of self-interference emerges as soon as the coherences become distributed across the entire spectrum and the initially Gaussian state completely delocalizes in spin space. The two snapshots on the right show a relatively short evolution time, where $\hat\rho_t$ is a spin-squeezed near-Gaussian state, and a longer time, where the state becomes clearly non-Gaussian and the $I_m$ develop an intricate structure for both the pure Ising and the transverse-field Ising case. This snapshot corresponds to the longest time that has been measured experimentally for these parameters in \cite{Gaerttner2016}. At this time, the $I_m$ fall off at most linearly with $m$, while the bound decreases exponentially [cf.\ Eq.~\eqref{eq:bound}]. This means that the degree ($I_m/I_m^{\rm max, sep}$) to which the entanglement bound is violated increases exponentially with $m$.

%
Next, we discuss the impact of decoherence for an example relevant to recent trapped-ion experiments~\cite{Bohnet2016,Gaerttner2016}. We find that decoherence can substantially reduce the state overlap $F_t(\phi)$. However, for the parameters of Ref.~\cite{Gaerttner2016}, detecting entanglement should be feasible. The main source of decoherence in these experiments is off-resonant light scattering, which can be captured by including Lindblad terms in the master equation \cite{SM}. Specifically, we consider elastic Rayleigh scattering, which leads to coherence decay with rate $\Gamma_{el}$, and Raman scattering, i.e., incoherent transitions from $\ket{\sd}$ to $\ket{\su}$ ($\Gamma_{du}$) and vice versa ($\Gamma_{ud}$).
We emphasize that if $\Gamma_{du}=\Gamma_{ud}$, which is typically the case in the trapped-ion experiments,  $\tr[\hat\rho_0\hat\rho_f]= \tr[\hat\rho_t \hat W(\phi) \hat\rho_t \hat W^\dagger(\phi)]$ in Eq.~\eqref{eq:scheme} still holds, and thus the $I_m$ can still be detected using the time reversal scheme \cite{SM}.

\begin{figure}[t]
  \centering
 \includegraphics[width=\columnwidth]{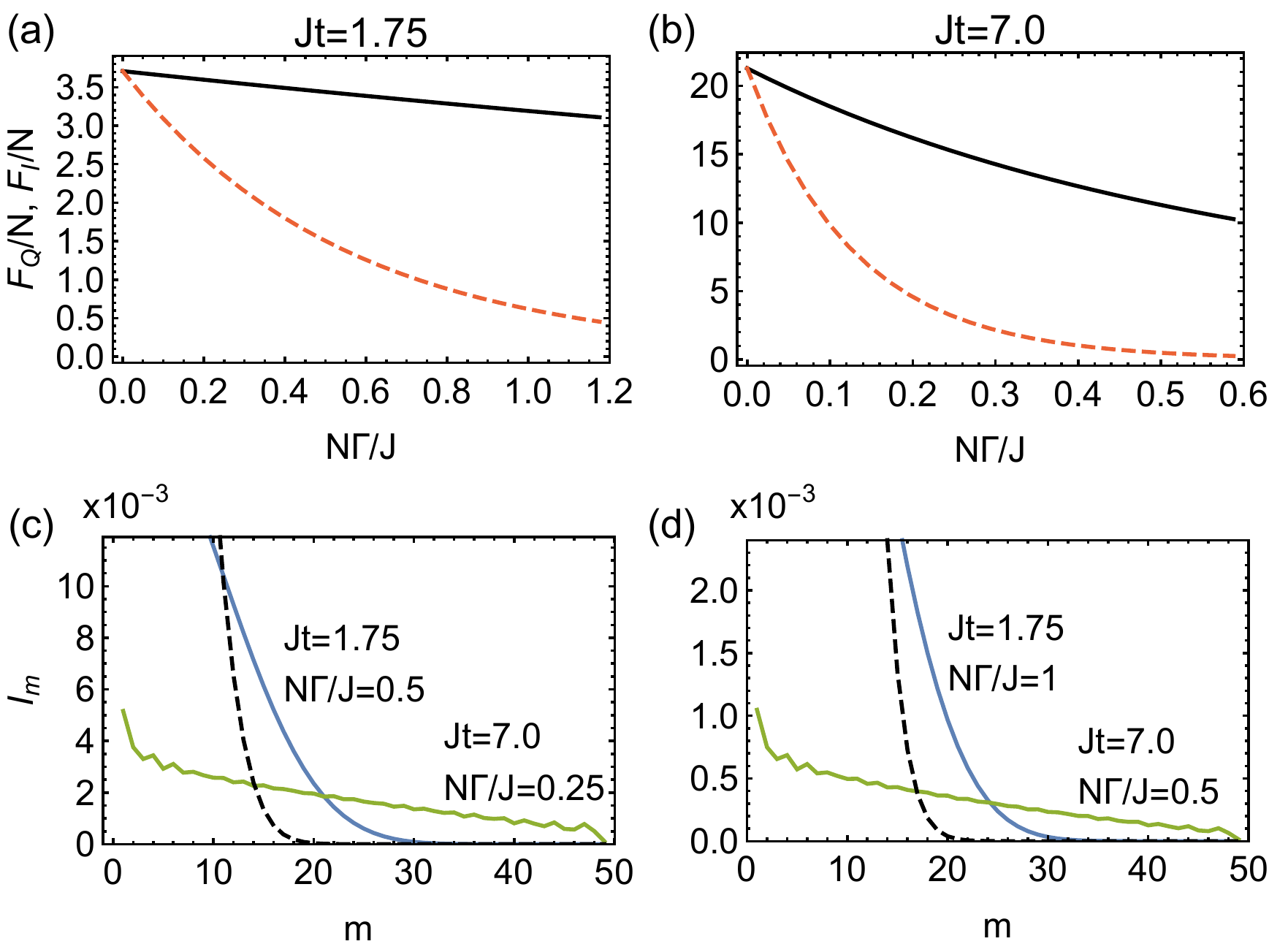}
\caption{(a) and (b) Optimal QFI (black) and the lower bound $F_I(t)$ (red dashed) as a function of the total decoherence rate $\Gamma=(\Gamma_{ud}+\Gamma_{du}+\Gamma_{el})/2$, scaled by $J/N$, and pure Ising dynamics ($\Omega=0$). The relative size of the individual decoherence rates for spontaneous emission and elastic scattering have been chosen $\Gamma_{ud}:\Gamma_{du}:\Gamma_{el}=1:1:10$. In the pure case ($\Gamma=0$) the bound coincides with the actual QFI. $F_I(\hat\rho_t,\hat A)$ decays as $\exp[-N\Gamma t]$, much faster than the QFI. 
The parameter choices are motivated by the parameters of Ref.~\cite{Gaerttner2016}, which corresponds to typical values of (a) $J=2.9\,$kHz and $t=0.6\,$ms and (b) $J=5.8\,$kHz and $t=1.2\,$ms. $N=48$ spins have been used.
(c) and (d) Coherences $I_m$ for two different dephasing rates in each case. Increasing the incoherent processes by a factor of two [comparing (c) with (d)], the coherences globally decrease but a violation of the entanglement bounds (dashed) is still found at large $m$.
For all values of $\Gamma$ the QFI is calculated with respect to the rotation axis $\mathbf{n}$ that is optimal for $\Gamma=0$.
}
\label{fig:deph}
\end{figure}

The role of decoherence is illustrated in Fig.~\ref{fig:deph}, where the choice of parameters is motivated by the experimental capabilities demonstrated in Ref.~\cite{Gaerttner2016}. Typical experimental parameters are $J\lesssim 5\,$kHz, $t\lesssim 1\,$ms, and a total decoherence rate $\Gamma\approx 60\,$s$^{-1}$, dominated by $\Gamma_{el}$. Numerical simulations were performed using an efficient density matrix symmetrization approach \cite{SM}. 

Comparing $F_Q(\hat\rho_t,\hat A)/N$ (black line) with the bound $F_I(\hat\rho_t,\hat A)/N$ (red dashed), one recovers $F_Q(\hat\rho_t,\hat A)=F_I(\hat\rho_t,\hat A)$ for pure states ($\Gamma=0$), but as decoherence rates are increased, the bound quickly becomes less tight. While the QFI decays slowly at small $\Gamma t$, the decay of the bound $F_I\sim e^{-N\Gamma t}$ is $N$-fold enhanced compared to the single-particle decay rate $\Gamma$ because the global state overlap $\tr[\hat\rho_0\hat\rho_f]$ decays with this rate. The inverse spin-squeezing parameter \cite{Kitagawa1993}, which also provides a lower bound on QFI, does not witness any entanglement for the case of Fig.~\ref{fig:deph}(b), as the state is already strongly oversqueezed.

Figures \ref{fig:deph}(c) and \ref{fig:deph}(d) show the coherence spectra for two values of $\Gamma$. The main effect of dephasing is a global decay of the $I_m$ with $e^{-N\Gamma t}$, approximately independent of $m$, as expected at short times in an initially pure system. Nevertheless, even for strong dephasing, the $I_m$ still violate the entanglement bound for sufficiently large $m$, since the bound decreases exponentially with $m$, while the $I_m$ decay much more slowly. 
Therefore, even in the presence of single-particle decoherence processes, we observe that the $I_m$ remain useful entanglement witnesses in the considered scenario. Nevertheless, one needs to deal with the experimental challenge of detecting a small signal, especially for large $N$. We note, however, that in Ref.~\cite{Gaerttner2016}, MQCs below $10^{-2}$ have been resolved.

%
In summary, we have derived inseparability criteria from the MQCs and a formal connection between MQCs and the QFI. Our results demonstrate that MQCs, a specific type of OTOCs, can serve as an experimentally accessible probe for detecting scrambling of quantum information and multiparticle entanglement in mixed states.

We thank
Arghavan Safavi-Naini, 
Michael Wall, 
John Bollinger,
Justin Bohnet,
Graeme Smith,
and Felix Leditzky
for discussions.
Supported by Defense Advanced Research Projects Agency
(DARPA, W911NF-16-1-0576 through ARO), NSF Grant No.\ PHY 1521080, JILA-NSF Grant No.\ NSF-PFC-PHY-1734006, 
AFOSR-MURI, NIST,
the DFG Collaborative Research Center SFB1225 (ISOQUANT),
the Austrian  Science  Fund  (FWF),  through SFB  FoQuS  (No.\  F4016-N23), the ERC Synergy Grant UQUAM, and the ERC Advanced Grant EntangleGen (Project-ID 694561).

\bibliographystyle{apsrev4-1}
\bibliography{MQC_theo}

\vspace*{1cm}

\clearpage
\newpage

\begin{widetext}

{\centering\textbf{\Large Supplemental Materials}}
\vspace{1cm}

In these supplementary materials, we describe the general scheme to experimentally extract multiple quantum coherences (MQCs) (Sec.~\ref{sec:ExtractingMQCs}) and discuss in which situations it remains valid even in the presence of decoherence (Sec.~\ref{sec:decoherence}). 
Moreover, we prove that the MQCs provide a lower bound on the quantum Fisher information (Sec.~\ref{sec:lowerboundQFI}). Further, we discuss the relation of MQCs to measures for coherence proposed in the literature and demonstrate some useful properties such as convexity and behavior under tensoring (Sec.~\ref{sec:propertiesOfMQC}), and we show that the individual MQCs are entanglement witnesses (Sec.~\ref{sec:entanglementBoundsForIm}). Finally, we provide technical details on the numerical calculation of the time evolution under the one-axis Hamiltonian with decoherence and the evaluation of relevant observables, exploiting the symmetry of the problem under particle exchange (Sec.~\ref{sec:numericalMethod}). Hats on operators are dropped here.

\section{Extracting MQCs from the many-body echo sequence\label{sec:ExtractingMQCs}}

In the time reversal sequence described in the main text, an initial state $\rho_0=\rho(t=0)$ evolves unitarily for a time $t$ into $\rho=e^{-iH\ind{int}t}\rho_0 e^{iH\ind{int}t}$. Then, the rotation $U(\phi)=\exp(iA\phi)$ is applied transforming $\rho$ into $\rho(\phi)=U(\phi)\rho U^\dagger(\phi)$, and subsequently the time evolution is reversed to give the final state $\rho_f=e^{iH\ind{int}t}\rho(\phi) e^{-iH\ind{int}t}$ 
Measuring the projector on $\rho(0)$ amounts to
\begin{equation}
  \begin{aligned}
  \tr[\rho_0\rho_f] & = \tr[\rho_0 e^{iH\ind{int}t} U(\phi) e^{-iH\ind{int}t} \rho_0 e^{iH\ind{int}t} U^\dagger(\phi) e^{-iH\ind{int}t}]\\
  & = \tr[e^{-iH\ind{int}t}\rho_0 e^{iH\ind{int}t} U(\phi) e^{-iH\ind{int}t} \rho_0 e^{iH\ind{int}t} U^\dagger(\phi)] \\
  & = \tr[\rho U(\phi) \rho U^\dagger(\phi)]  = \tr[\rho\rho(\phi)] \\
  & = \tr[\sum_{m^\prime} \rho_{m^\prime} \sum_m \rho_m e^{-i m\phi} ] \\
  & = \sum_m \tr[\rho_{-m}\rho_m] e^{-i m\phi} \equiv \sum_m I_m(\rho) e^{-i m\phi}
  \label{eq:scheme2}
  \end{aligned}
 \end{equation}
Thus, the measured observable is the Fourier transform of the multiple-quantum coherence spectrum of the state $\rho$ and the MQCs $I_m$ can be extracted from it.

In the simulations presented in the main text, we focus on ensembles of $N$ (pseudo-) spin $1/2$ particles and consider coherences with respect to the generator of a global rotation $A=S_\mathbf{n} = n_x S_x + n_y S_y + n_z S_z$. In this case, the spectrum of $A$ consists of (half) integer values $-N/2\ldots N/2$, and thus $m=-N\ldots N$ and all coherences can be extracted by scanning the phase in $U(\phi)$ over the range $[0,2\pi]$. Generalizations to global rotations in higher spin systems are straight forward. Also, for $A=1/2\sum_j (n_j^x \sigma_j^x + n_j^y \sigma_j^y + n_j^z \sigma_j^z)$, i.e.\ rotating each spin about an individual axis, none of our conclusions change.

\section{Effect of decoherence on the MQC detection scheme\label{sec:decoherence}}

The above equations assumed a unitary evolution. However, realistic experiments will suffer from diverse sources of decoherence. While these may suppress the MQCs, as discussed in the main text, we now show that the scheme for detecting the MQCs works equally well in a range of situations where decoherence is present. 

Assuming decoherence due to a featureless Markovian bath, as is usually the case in trapped-ion or ultracold-atoms experiments, we can take its effect into account through a Lindblad master equation 
\begin{equation}
\label{eq:ME}
\dot{\rho}=-i[H\ind{int},\rho]+\sum_n \mathcal{L}(\hat{\Gamma}_n)\rho\,,
\end{equation}
where 
\begin{equation}
\mathcal{L}(\hat{\Gamma}_n)\rho = \sum_j\hat{\Gamma}_{n,j}\rho \hat{\Gamma}_{n,j}^\dagger  -\frac{1}{2}(\hat{\Gamma}_{n,j}^\dagger \hat{\Gamma}_{n,j}\rho+ \rho\hat{\Gamma}_{n,j}^\dagger \hat{\Gamma}_{n,j} )
\end{equation}
is a Lindblad operator ($j$ is the particle index). Note, that here we use hats on the jump operators $\hat \Gamma$ to distinguish them from their corresponding jump rates $\Gamma$. The most relevant types of decoherence are spontaneous emission up ($\hat{\Gamma}_{du}=\sqrt{\Gamma_{du}}\ket{\su}\bra{\sd}$) and down ($\hat{\Gamma}_{ud}=\sqrt{\Gamma_{ud}}\ket{\sd}\bra{\su}$) as well as elastic dephasing ($\hat{\Gamma}_{el}=\sqrt{\Gamma_{el}}\ket{\su}\bra{\su}$, or, more generally, $\hat{\Gamma}_{el}^\mathbf{n}=\sqrt{\Gamma_{el}}\ket{\su}_\mathbf{n}\bra{\su}_\mathbf{n}$, where $\ket{\su}_\mathbf{n}$ is the up-eigenstate in the $\mathbf{n}$ direction on the Bloch sphere). 
Often, one faces situations where the dominant contribution is either only from elastic dephasing \cite{Uys2010} or where $\Gamma_{du}\approx \Gamma_{ud}$ \cite{Schindler2013}. In both cases, the detection scheme remains valid, as we show now. 

What we need to prove is that 
\begin{equation}
\tr[\rho_0\rho_f]=\tr[\rho_t\rho_t(\phi)]
\end{equation}
or
\begin{equation}
\label{eq:expdissipationback}
\tr[\rho_0 (\exp[\mathcal{L}\ind{back}t]\rho_t(\phi))]=\tr[(\exp[\mathcal{L}t]\rho_0)\rho_t(\phi)]
\end{equation}
where $\mathcal{L}\rho=-i[H,\rho]+\mathcal{L}\ind{diss}[\rho]$ and $\mathcal{L}\ind{back}\rho=-i[-H,\rho]+\mathcal{L}\ind{diss}[\rho]$.
By writing the time evolution as a Trotter expansion, we only have to prove the above for an infinitesimal time step, which can then be applied iteratively.
Expanding the exponential for an infinitesimal step $\Delta t$, we have
\begin{equation}
\begin{aligned}
\tr[\rho_0 \exp[\mathcal{L}\ind{back}\Delta t]\rho_t(\phi)] &= \tr[(\exp[\mathcal{L}\Delta t]\rho_0)\rho_t(\phi)] \\
\Leftrightarrow \tr[\rho_0(\mathcal{L}\ind{back}\rho_t(\phi))] &=\tr[(\mathcal{L}\rho_0) \rho_t(\phi)]\,.
\end{aligned}
\end{equation}
Using $\tr(\rho_0[H,\rho_t(\phi)])=-\tr([H,\rho_0]\rho_t(\phi))$ due to cyclic permutation under the trace, what remains to be shown is 
\begin{equation}
\label{eq:dissipation_back}
\tr[\rho_0(\mathcal{L}\ind{diss}\rho_t(\phi))] =\tr[(\mathcal{L}\ind{diss}\rho_0) \rho_t(\phi)]\,.
\end{equation}
For this, we note $\tr[\rho_1(\mathcal{L}(\hat{\Gamma})\rho_2)]=\tr[(\mathcal{L}(\hat{\Gamma}^\dagger)\rho_1)\rho_2]$. 
Using this relation together with  $\mathcal{L}\ind{diss}\rho=(\mathcal{L}[\hat{\Gamma}_{el}]+\mathcal{L}[\hat{\Gamma}_{ud}]+\mathcal{L}[\hat{\Gamma}_{ud}^\dagger])\rho$, valid for  $\Gamma_{ud}=\Gamma_{du}$, and the fact that $\hat{\Gamma}_{el}=\hat{\Gamma}_{el}^\dagger$, demonstrates Eq.~\eqref{eq:dissipation_back} and in consequence also Eq.~\eqref{eq:expdissipationback}.

\section{Derivation of the lower bound on quantum Fisher information\label{sec:lowerboundQFI}}

The statement $4\tr[\rho^2 A^2-(\rho A)^2]=2\sum_m I_m(\rho) m^2$ used in the main text (time arguments have been dropped) can be proven by expanding $\tr[\rho\rho(\phi)]$ in $\phi$ around $\phi=0$
\begin{equation}
\begin{aligned}
    \tr[\rho\,\rho(\phi)] &= \tr[\rho e^{-iA\phi}\rho e^{iA\phi}]  \\
  & = \tr[\rho(1-iA\phi-\frac{1}{2}A^2\phi^2+\mathcal{O}(\phi^3)) \rho (1+iA\phi-\frac{1}{2}A^2\phi^2+\mathcal{O}(\phi^3))] \\
  & = \tr[\rho^2]-\phi^2 \tr[\rho^2 A^2-(\rho A)^2] +\mathcal{O}(\phi^3)
\end{aligned}
\end{equation}
Taking the second derivative with respect to $\phi$ this gives
\begin{equation}
\begin{aligned}
    2\tr[\rho^2 A^2-(\rho A)^2] & = \left. -\frac{d^2}{d\phi^2}\tr[\rho\rho(\phi)]\right|_{\phi=0} \\
    & = \left. -\frac{d^2}{d\phi^2}\sum_m I_m(\rho) e^{-im\phi}\right|_{\phi=0} \\
    &= \sum_m m^2 I_m(\rho) \,.
\end{aligned}
\end{equation}
The inequality $F_Q(\rho,A) \geq 4\tr[\rho^2 A^2-(\rho A)^2]$ \cite{Girolami2015} follows from the relation of the QFI with the Uhlmann fidelity $f(\rho,\rho(\phi))$ \cite{Jozsa1994, Liu2014}
\begin{equation}
\begin{aligned}
F_Q(\rho,A) &=\left.-2\frac{d^2}{d\phi^2} f(\rho,\rho(\phi))\right|_{\phi=0} \\
&=\left.-2\frac{d^2}{d\phi^2} \left(\tr\left[\sqrt{\sqrt{\rho}\rho(\phi)\sqrt{\rho}}\right]\right)^2\right|_{\phi=0}
\end{aligned}
\end{equation}
and the inequality \cite{Miszczak2009}
\begin{equation}
\begin{aligned}
    f(\rho,\rho(\phi))&\geq \tr[\rho\,\rho(\phi)] + \sqrt{(1-\tr[\rho^2])(1-\tr[\rho(\phi)^2])} \\
   & = \tr[\rho\,\rho(\phi)] + 1-\tr[\rho^2] \,.
\end{aligned}
\end{equation}

\section{Properties of the MQCs\label{sec:propertiesOfMQC}}

{\bf Classification of MQCs in common categories of coherence measures:} In the main text, we define the multiple quantum coherences of a quantum state $\rho$ with respect to a hermitian operator $A$ as $\rho_m=\sum_{M-M^\prime=m} \rho_{M^\prime M}\ket{M^\prime}\bra{M}$, where $\ket{M}$ are the eigenstates of $A$ with eigenvalue $M$. The $\rho_m$ are identical with the $\delta$-coherences defined in Ref.~\cite{Yadin2016}, which quantify the macroscopicity of a state \cite{Kwon2015}. 
The function $F_I(\rho,A)=\sum_m m^2 I_m(\rho)$ defined in the main text as well as the quantum Fisher information are examples for measures of $\delta$-coherence. 
The $\delta$-coherence is a special case of the translationally covariant (TC) coherence defined in Ref.~\cite{MarvianSpekkens2016}. In that definition, a state $\rho=\sum_m \rho_m$ is called incoherent with respect to $A$ if it is invariant under the action of the unitary generated by $A$, i.e., if $e^{-iA\phi}\rho e^{iA\phi}=\sum_m\rho_m e^{-im\phi}=\rho$. This is the case if and only if $\rho_m=0\quad \forall\, m\neq 0$. 
The generalized rotation $U(\phi)$ generated by $A$ is called a translation, whence the TC coherence takes its name. The trace norm $\lvert\rho_m\rvert_1 = \tr[\sqrt{\rho_m^\dagger \rho_m}]$, \cite{MarvianSpekkens2016} as well as the quantum Fisher information \cite{Girolami2014} are examples for measures of TC-coherence. 

These definitions quantify the type of coherence that is relevant for tasks appearing, e.g., in quantum metrology where a distinguished operator $A$ exists. For example, in quantum phase estimation $A$ generates translations of a phase that one seeks to determine. For such tasks, off-diagonal entries in the density matrix that connect states with the same eigenvalue of $A$ do not constitute relevant coherences, as they are insensitive towards the translations generated by $A$ \cite{MarvianSpekkens2016}. 
Except in the case where the spectrum of $A$ is non-degenerate, these definitions deviate from the one of Ref.~\cite{Baumgratz2014}, which considers all off-diagonal entries of the density matrix as coherent. 

{\bf Convexity:} The MQCs of a mixture of two (or more) states ($\rho=\sum_k p_k\rho_k$, where $\sum_k p_k=1$) cannot become larger than the weighted sum of the MQCs of the components of the mixture: 
\begin{equation}
 I_m\left(\sum_k p_k\rho_k \right) \leq \sum_k p_k I_m(\rho_k)\,.
\end{equation}
\\
\textit{Proof:}
We prove the inequality for the special case $\rho=p\rho^{(1)}+(1-p)\rho^{(2)}$, which can be immediately generalized to arbitrary $\rho=\sum_k p_k\rho_k$.  For $p=0$ or $p=1$ the inequality follows trivially, so we need only prove the case $p\neq 0,1$. 
\begin{equation}
 \begin{aligned}
  I_m(p\rho^{(1)}+(1-p)\rho^{(2)}) = \tr[(p\rho_{-m}^{(1)}+(1-p)\rho_{-m}^{(2)})(p\rho_{m}^{(1)}+(1-p)\rho_{m}^{(2)})] & \leq  p\tr[\rho_{-m}^{(1)}\rho_{m}^{(1)}]+ (1-p)\tr[\rho_{-m}^{(2)}\rho_{m}^{(2)}] \\
   \Leftrightarrow  p^2 \tr[\rho_{-m}^{(1)}\rho_{m}^{(1)}] + (1-p)^2\tr[\rho_{-m}^{(2)}\rho_{m}^{(2)}] + p(1-p)  \tr[\rho_{-m}^{(1)}\rho_{m}^{(2)}+\rho_{-m}^{(2)}\rho_{m}^{(1)}] & \leq  p\tr[\rho_{-m}^{(1)}\rho_{m}^{(1)}] +(1-p)\tr[\rho_{-m}^{(2)}\rho_{m}^{(2)}] \\
   \Leftrightarrow p(1-p)\tr[\rho_{-m}^{(1)}\rho_{m}^{(1)} + \rho_{-m}^{(2)}\rho_{m}^{(2)} -\rho_{-m}^{(1)}\rho_{m}^{(2)}-\rho_{-m}^{(2)}\rho_{m}^{(1)}] & \geq 0 \\
   \Leftrightarrow \tr[(\rho_{-m}^{(1)}-\rho_{-m}^{(2)})(\rho_{m}^{(1)}-\rho_{m}^{(2)})] & \geq 0 \\
   \Leftrightarrow \tr[(\Delta\rho_{m}^\dagger \Delta\rho_{m}] & \geq 0 \,,
 \end{aligned}
\end{equation}
where the last statement is true since the left side is just the (non-negative) Frobenius norm of $\Delta\rho_{m}$.

{\bf Direct product:} We want to express the MQCs of a state $\rho$ that is a direct product of two subsystems $\rho=\rho_A\otimes \rho_B$ by the MQCs of $\rho_A$ and $\rho_B$. Collecting the $m$th order coherences of $\rho=\rho_A\otimes \rho_B$ (writing $\rho_X=\sum_m \rho_m^{X}$) we have to sum over all the tensor products of sectors such that the sum of the coherenes is $m$: $\rho_m=\sum_k \rho_{m-k}^{A}\otimes \rho_{m+k}^{B}$. In the case of a state of $N_A+N_B$ spin $1/2$ particles and $U(\phi)$ a global rotation the sum runs over indices such that $m-k\in[-N_A,N_A]$ and $m+k\in [-N_B,N_B]$. Thus for the $I_m$ we obtain 
\begin{equation}
\begin{aligned}
 I_m(\rho) &= \tr[\rho_{-m}\rho_m] \\
     &= \tr[(\sum_k \rho_{-m-k}^A\otimes\rho_{-m+k}^B)(\sum_{k^\prime} \rho_{m-k^\prime}^A\otimes\rho_{m+k^\prime}^B)] \\
     &=\sum_{kk^\prime} \tr[(\rho_{-m-k}^A\rho_{m-k^\prime}^A)\otimes(\rho_{-m+k}^B\rho_{m+k^\prime}^B)] \\
     &=\sum_{k} \tr[(\rho_{-m-k}^A\rho_{m+k}^A)\otimes(\rho_{-m+k}^B\rho_{m-k}^B)] \\
     &=\sum_k  I_{m+k}(\rho_A)I_{m-k}(\rho_B)\,,
\end{aligned}
\end{equation}
where we used the distributive law of inner and outer products, $\tr[A\otimes B]=\tr[A]\tr[B]$, and the fact that the $\tr[\rho_m\rho_{m^\prime}]\neq 0$ only if $m=-m^\prime$.

\section{Deriving entanglement bounds for the individual $I_m$\label{sec:entanglementBoundsForIm}}

Next, we sketch the proof for the entanglement bounds on the individual coherences $I_m$ stated in the main text.
We first calculate the MQCs of a coherent spin state (CSS). A general CSS can be expressed in terms of fully symmetric Dicke states. 
\begin{equation}
|\text{CSS}(\theta,\phi)\rangle = \left( \sin(\theta/2)\ket{\su}+e^{i\varphi}\cos(\theta/2)\ket{\sd}\right)^{\otimes N} =\sum_{k=0}^N\alpha_{N,k}|N,k\rangle\,,
\end{equation}
where
\begin{equation}
    \alpha_{N,k} = \sqrt{ \binom{N}{k}}\sin(\theta/2)^{N-k}\cos(\theta/2)^k e^{ik\varphi}
\end{equation}
and
\begin{equation}
    |N,k\rangle = \sqrt{\frac{(N-k)!}{N!k!}} \left(\sum_{i=1}^N \ket{\su_i}\bra{\sd_i}\right)^k \ket{\su\ldots \su}
\end{equation}
are the symmetric Dicke states.

Without loss of generality (see also Sec.~\ref{sec:ExtractingMQCs}), we restrict ourselves to the coherences with respect to the basis $\{\ket{\su},\ket{\sd}\}$, which usually denotes the eigenstates of $\sigma_z$, but can be the eigenstates of any spin operator $\mathbf{n}\cdot \boldsymbol{\sigma}$.
Thus, we obtain for the CSS
\begin{equation}
 \rho_m=\sum_{k=0}^{N-m}\alpha_{N,k}\alpha^*_{N,k+m}|N,k\rangle\langle N,k+m|=\rho_{-m}^\dagger
\end{equation}
and thus the coherence spectrum of the CSS with $\theta=\pi/2$ is
\begin{equation}
\begin{aligned}
 I_m(\ket{\rm CSS}\bra{\rm CSS}) &= \sum_{k=0}^{N-m}|\alpha_{N,k}|^2 |\alpha_{N,k+m}|^2 \\
 & = \frac{(2N)!}{4^N(N-m)!(N+m)!}\,.
\end{aligned}
\end{equation}

We now proceed to derive entanglement bounds for the $I_m$. For this, we maximize $I_m$ on the set of fully separable states
\begin{equation}
 \rho\ind{sep}=\sum_k \beta_k \bigotimes_j \rho_{k}^{(j)}
\end{equation}
where $\sum_k \beta_k=1$ and $\rho_{k}^{(j)}$ general single particle densities. The convexity of the $I_m$ yields
\begin{equation}
 I_m(\rho\ind{sep})\leq \sum_k \beta_k I_m(\bigotimes_j \rho_{k}^{(j)})\leq \max_k I_m(\bigotimes_j \rho_{k}^{(j)})
\end{equation}
So the supremum of $I_m$ can be upper bounded by optimizing over pure product states
\begin{equation}
    \ket{\psi\ind{sep}}=\bigotimes_{j=1}^N \ket{\psi_j}=\bigotimes_{j=1}^N (\sqrt{p_j}\ket{\su}+e^{i\varphi_j}\sqrt{1-p_j}\ket{\sd})
\end{equation}
Using their behavior under tensoring, the coherences $I_m$ can be calculated recursively from the coherences of the individual spin states $\rho_j=\ket{\psi_j}\bra{\psi_j}$:
\begin{equation}
\label{eq:ImRecursion}
 I_m(\rho_1 \otimes \ldots \rho_N)=I_{m-1}(\rho_1\otimes \ldots \rho_{N-1})I_{1}(\rho_N)+I_{m}(\rho_1\otimes \ldots \rho_{N-1})I_{0}(\rho_N)+I_{m+1}(\rho_1\otimes \ldots \rho_{N-1})I_{-1}(\rho_N)
\end{equation}
The single-particle MQCs are
\begin{equation}
 \begin{aligned}
  I_0(\rho_j) &= p_j^2+(1-p_j)^2 \\
  I_1(\rho_j) = I_{-1}(\rho_j) &= p_j(1-p_j)
 \end{aligned}
\end{equation}
Thus $I_m$ is a second order polynomial in $p_j$. We also notice that $I_0(\rho_j)$ and $I_1(\rho_j)$ (and thus $I_m$) are invariant under the substitution $p_j\rightarrow 1-p_j$. Thus the maximum with respect to $p_j$ is assumed either at $p_j=1/2$ (if $d^2I_m/dp_j^2 <0$) or at the edges $p_j=0,1$. Hence, in order to maximize $I_m$, each spin must either be in the state $(\ket{\su}+\ket{\sd})/\sqrt{2}$ or in one of $\ket{\su}$ and $\ket{\sd}$. According to equation~\eqref{eq:ImRecursion}, $I_m$ does not change if a particle in $\ket{\su}$ or $\ket{\sd}$ is added. Thus, the maximal $I_m$ for a separable state of $N$ spins equals the coherence of a coherent spin state of $N\ind{opt}$ particles with $\theta=\pi/2$, where $N\ind{opt}\in \{0,1\ldots N\}$ maximizes $I_m$.

\section{Numerical method\label{sec:numericalMethod}}

\subsection{Symmetrized Liouville space}

To study the MQCs numerically, we implemented the master equation \eqref{eq:ME} with $H$ the one-axis twisting Hamiltonian and with Lindblad operators $\hat{\Gamma}_{du}$, $\hat{\Gamma}_{ud}$, and $\hat{\Gamma}_{el}$. 
As long as the interactions are all-to-all and the decoherence processes affect all particles in the same way, Eq.~\eqref{eq:ME} is invariant under exchange of  particles. Since the global rotations involved in the sequence also preserve this symmetry and since the initial state is fully symmetric under particle exchange, the dynamics is restricted to the space of fully symmetric density matrices \cite{Sarkar1987, Hartmann2012, Xu2013}, allowing for an efficient numerical implementation. 

In general, any density matrix can be represented as
\begin{equation}
 \rho=\sum_\alpha c_\alpha \rho_\alpha\,,
\end{equation}
where 
\begin{equation}
 \rho_\alpha=\bigotimes_{j=1}^N \sigma_{\alpha_j}
\end{equation}
and $\sigma_{\alpha_j} \in \{1,\sigma_z,\sigma_+,\sigma_-\}$. Here, $\alpha$ represents the vector of the single particle states for a given basis state.
The symmetry constraint now means that this Liouville space basis can be restricted to symmetrized states
\begin{equation}
 (n_z,n_+,n_-)=\mathcal{N}^{-1}\sum_{\chi\in S_N} \rho_{\chi(\alpha)}\,,
\end{equation}
where $(n_z,n_+,n_-)$ are the number of occurrences of $(z,+,-)$ in $\alpha$ (the index $1$ occurs $n_1=N-n_z-n_+-n_-$ times) and $\chi(\alpha)$ is a permutations of the indices. There are $N!$ permutations but many of them generate identical states. We choose the normalization factor $\mathcal{N}=n_z!n_+!n_-!n_1!$ such that all (different) states occur with unit weight. Thus the possible symmetric states are effectively given by the Fock states of $N$ bosons on a 4-site lattice, which gives a dimension of
\begin{equation}
 d\ind{sym}=\binom{N+3}{3}\sim \frac{N^3}{6}\,,
\end{equation}
which is a tremendous reduction of dimensionality compared to the $4^N$ basis states needed to represent a general $N$-particle density matrix. We will later make use of other basis choices, like $\sigma_{\alpha_j} \in \{1,\sigma_x,\sigma_y,\sigma_z\}$, but in the context of the MQC spectrum the $+/-$ basis is very intuitive since the block structure in terms of coherence blocks $m=n_+ - n_-$ is already built in.

In order to carry out calculations on the symmetrized subspace we only have to write the master equation as well as the initial state and the observables in terms of the coefficients $c_\alpha$.

\begin{itemize}

 \item {\bf Initial state:} 
 \begin{equation}
  \bigotimes_j\ket{\su_j}\bra{\su_j} = \bigotimes_j (1+\sigma_z^{(j)})/2 = \sum_{n_z=0}^N \frac{1}{2^N}(n_z,0,0)
 \end{equation}
 
 \item {\bf Liouvillian:} In order to calculate the matrix elements of the relevant Liouvillian operator we calculate its action on each basis state $(n_z,n_+,n_-)$, i.e., we determine the coefficients $a_{\alpha\beta}$ in $\mathcal{L}[\rho_\alpha]=\sum_{\beta}a_{\alpha\beta}\rho_\beta$

\emph{Interactions:} 
  \begin{equation}
   -i[\sum_{k<j}\sigma_z^{(k)}\sigma_z^{(j)},(n_z,n_+,n_-)]=-2i(n_+-n_-)[(n_z+1)(n_z+1,n_+,n_-)+(n_1+1)(n_z-1,n_+,n_-)]
  \end{equation}
  This term conserves $n_+$ and $n_-$ and is thus block diagonal if written in blocks of fixed $(n_+,n_-)$. These blocks are tri-diagonal.

\emph{Incoherent terms:} We consider the Linbald terms, e.g.,
  \begin{equation}
   \mathcal{L}\ind{ud}[\rho]=\Gamma\ind{ud}\sum_j \sigma_-^{(j)} \rho \sigma_+^{(j)} -\frac{1}{2}[ \sigma_+^{(j)}\sigma_-^{(j)} \rho + \rho \sigma_+^{(j)}\sigma_-^{(j)} ]\,,
  \end{equation}
  which gives
  \begin{equation}
   \mathcal{L}\ind{ud}[(n_z,n_+,n_-)] = -\Gamma\ind{ud}\{ (n_z+1)(n_z+1,n_+,n_-) + [n_z+(n_++n_-)/2](n_z,n_+,n_-)\}\,,
  \end{equation}
  \begin{equation}
   \mathcal{L}\ind{du}[(n_z,n_+,n_-)] = -\Gamma\ind{du}\{ -(n_z+1)(n_z+1,n_+,n_-) + [n_z+(n_++n_-)/2](n_z,n_+,n_-)\}\,,
  \end{equation}
  \begin{equation}
   \mathcal{L}\ind{el}[(n_z,n_+,n_-)] = -\Gamma\ind{el}(n_++n_-)/2(n_z,n_+,n_-)\,.
  \end{equation}

 Here, we can see that $n_+$ and $n_-$ are conserved and if $\Gamma\ind{ud}=\Gamma\ind{du}$, the whole dissipative term is diagonal, i.e., just leads to an exponential decay of coherences. The joint action of one-axis twisting and dissipation is block diagonal with block size at most $N+1$, which greatly simplifies the problem.
 
 With the above, the master equation can be written in terms of the coefficients $c_\alpha$:
 \begin{equation}
  \begin{aligned}
   \dot{\rho} &= \mathcal{L}[\rho] \\
   \sum_{\alpha^\prime} \dot{c}_{\alpha^\prime} \rho_{\alpha^\prime} &= \mathcal{L}[\sum_{\alpha^\prime} c_{\alpha^\prime} \rho_{\alpha^\prime} ] =  \sum_{\alpha^\prime} c_{\alpha^\prime} \mathcal{L}[\rho_{\alpha^\prime} ]=  \sum_{\alpha^\prime} c_{\alpha^\prime} \sum_\beta a_{\alpha^\prime\beta}\rho_\beta \\
   \sum_{\alpha^\prime} \dot{c}_{\alpha^\prime} \tr[\rho_{\alpha^\prime}\rho_{\alpha}^\dagger] &= \sum_{\alpha^\prime} c_{\alpha^\prime} \sum_\beta a_{\alpha^\prime\beta} \tr[\rho_\beta\rho_\alpha^\dagger] \\
   \sum_{\alpha^\prime} \dot{c}_{\alpha^\prime} \tr[\rho_{\alpha}\rho_{\alpha}^\dagger]\delta_{\alpha\alpha^\prime} &= \sum_{\alpha^\prime} c_{\alpha^\prime} \sum_\beta a_{\alpha^\prime\beta} \tr[\rho_\alpha\rho_{\alpha}^\dagger]\delta_{\alpha\beta} \\
   \dot{c}_{\alpha} &= \sum_{\alpha^\prime} a_{\alpha^\prime\alpha}c_{\alpha^\prime} 
  \end{aligned}
 \end{equation}
 Thus the matrix-elements calculated above have to be transposed to yield the master equation for coefficient vector $c_\alpha$. We used that the basis states $\rho_\alpha$ are orthogonal, i.e. $\tr[\rho_\alpha\rho_\beta^\dagger]\propto \delta_{\alpha\beta}$. 
 
 \item {\bf Observables} are expressed in terms of the $c_\alpha$ by expressing them in terms of the basis states $\rho_\alpha$.\\
 We will repeatedly make use of
 \begin{equation}
  \tr[(n_z,n_+,n_-)(n_z,n_+,n_-)^\dagger]=2^{n_1+n_z}\frac{N!}{n_1!n_z!n_+!n_-!}\,.
 \end{equation}
 The factor $2^{n_1+n_z}$ is due the fact that $\tr[1^2]=\tr[\sigma_z^2]=2$, and could be removed by defining the basis in terms of $1/2$ and $\sigma_z/2$. Similarly, one could normalize the basis states by the number of permutations, but then those factors would appear in the initial state.\\ 
 Components of total spin (or populations $N_\su=N/2+\langle S_z \rangle$):
 \begin{equation}
  \langle S_z\rangle= \tr[ S_z\rho] = \tr[1/2(1,0,0)\sum c_\alpha \rho_\alpha]=c_{(1,0,0)}/2\tr[(1,0,0)^2] = c_{(1,0,0)}\frac{N}{2}2^N\,,
 \end{equation}
 \begin{equation}
  \langle S_x\rangle= \tr[ (S_+-S_-)\rho] = \tr[1/2[(0,1,0)+(0,0,1)]\sum c_\alpha \rho_\alpha]= \frac{N}{2}2^{N-1} (c_{(0,0,1)}+c_{(0,1,0)})\,,
 \end{equation}
 \begin{equation}
  \langle S_y\rangle= \tr[i(- S_+ +S_-)\rho] = \tr[i/2[-(0,1,0)+(0,0,1)]\sum c_\alpha \rho_\alpha]= i\frac{N}{2}2^{N-1} (-c_{(0,0,1)} + c_{(0,1,0)})\,.
 \end{equation}
 The all up probability: 
 \begin{equation}
  P_0=\langle \ket{\su\ldots \su}\bra{\su\ldots \su}\rangle= \tr[\sum_{n_z}(n_z,0,0)/2^N\rho] = \sum_{n_z}c_{(n_z,0,0)} \binom{N}{n_z}\,.
 \end{equation}
 Muliple quantum coherences:
 \begin{equation}
   \rho_m=\sum_{\alpha:n_+-n_-=m}c_\alpha \rho_\alpha\,.
 \end{equation}
 \begin{equation}
  I_m(\rho)=\tr[\rho_{-m}\rho_m]=\tr[\rho_{m}^\dagger\rho_m]=\sum_{n_z,n_+}|c_{(n_z,n_+,m-n_+)}|^2\frac{2^{n_1+n_z}N!}{n_1!n_z!n_+!(m-n_+)!}\,.
 \end{equation}
 Second moment of the collective spin:
 \begin{equation}
  \langle S_z^2\rangle = \left(\sigma_z^{(1)}/2+\ldots +\sigma_z^{(N)}/2 \right)^2=\frac{1}{4}\langle \sum_{k,j} \sigma_z^{(k)} \sigma_z^{(j)} \rangle =  \frac{1}{4}\langle N+\sum_{k\neq j} \sigma_z^{(k)} \sigma_z^{(j)} \rangle = \frac{N}{4} + \frac{2^N}{2}\binom{N}{2}c_{(2,0,0)}\,.
 \end{equation}
 $m$th moment:
 \begin{equation}
  \langle S_z^m\rangle=\sum_{n=0}^N (N/2-n)^m P_n\,, 
 \end{equation}
 where $P_n$ is the probability to find $n$ spins in state $\ket{\su}$. $P_n$ is easier to directly compute by combinatoric arguments than $\langle S_z^m\rangle$. To calculate the moments of the spin along other axes $S_{\mathbf{n}}^m$, one can simply rotate the state into the corresponding direction and then calculate $S_{z}^m$.

Probability $P_n$ for $n$ spins in $\ket{\su}$: 
 \begin{equation}
 \begin{aligned}
  P_n &=\sum_{n_z=0}^N c_{(n_z,0,0)} \binom{N}{n}\sum_{k=0}^n \binom{N-n}{n_z-k}\binom{n}{k}(-1)^k \\
  &= \sum_{n_z=0}^N c_{(n_z,0,0)} \binom{N}{n}\binom{N-n}{n_z} {_2}F_1(-n,-n_z,N-n-n_z+1;-1)\,,
 \end{aligned}
 \end{equation}
 where binomial coefficients with $n<k$ are defined to be zero and ${_2}F_1$ denotes a hypergeometric function.
\end{itemize}

\subsection{Efficient implementation}

We noticed that the the evolution under one-axis twisting and dissipation has a particularly simple form in the $(n_z,n_+,n_-)$ basis. The Liouvillian $\mathcal{L}\ind{int}$ is block-diagonal with blocks of size $\leq N+1$ and the blocks are tri-diagonal matrices. The map $\exp[\mathcal{L}\ind{int}t]$ can thus be evaluated efficiently, scaling as $N^2$ for each block, giving an overall scaling of $N^4$ (there are $\mathcal{O}[N^2]$ blocks). Similarly, rotations are block-diagonal with blocks of size $\leq N+1$ in the $(n_x,n_y,n_z)$ basis because, e.g., for a rotation about $y$, $n_y$ as well as $n_x+n_z$ is conserved. Moreover, the basis transformation between $(z+-)$ and $(xyz)$ basis also has such block structure since $n_z$ is conserved and $n_x+n_y=n_++n_-$. We can thus write a rotation in the $(z+-)$ basis as a sequence of simpler operations: Transform to $(xyz)$, rotate, transform back. Each of these steps is a matrix-vector multiplication where each matrix has $\mathcal{O}[N^4]$ non-zero elements (recall that a rotation in the $(z+-)$ basis has $\mathcal{O}[N^5]$ non-zero elements).

We compute the matrix elements for rotations in the $(n_x,n_y,n_z)$ basis by working out the action of $\exp[\mathcal{L}\ind{rot}\phi]$ on the basis states. Here,  we give the matrix elements for $y$-rotations ($S_y=\sum_i \sigma_y^{(i)}/2$),
\begin{equation}
 e^{-iS_y \phi}(n_x,n_y,n_z)e^{iS_y \phi}=\sum_{n_x^\prime=0}^{n_x+n_z}A_{n_x,n_z}^{n_x^\prime,n_z^\prime=n_x+n_z-n_x^\prime}(n_x^\prime,n_y,n_x+n_z-n_x^\prime)\,,
\end{equation}
with 
\begin{equation}
 A_{n_x,n_z}^{n_x^\prime,n_z^\prime=n_x+n_z-n_x^\prime}=\frac{n_x^\prime! n_z^\prime!}{n_x!n_z!}\sum_{n_{xz}=\max[0,n_x^\prime-n_z]}^{\min[n_x^\prime,n_x]}\binom{n_x}{n_{xz}}\binom{n_z}{n_x^\prime-n_{xz}}(\cos\phi)^{n_z-nx^\prime+2n_{xz}}(\sin\phi)^{n_x+n_x^\prime-2n_{xz}}(-1)^{n_x-n_{xz}}\,.
\end{equation}

The transformations between the two different basis sets is given by
\begin{equation}
 (n_z,n_+,n_-)=\sum_{n_x=0}^{n_++n_-}A_{n_+,n_-}^{n_x,n_y=n_++n_--n_x} (n_x,n_y,n_z)\,,
\end{equation}
with
\begin{equation}
 A_{n_+,n_-}^{n_x,n_y=n_++n_--n_x} = \frac{n_x! n_y!}{n_+!n_-!}\sum_{n_{x+}=\max[0,n_x-n_-]}^{\min[n_x,n_+]}\binom{n_+}{n_{x+}}\binom{n_-}{n_x-n_{x+}}\frac{i^{n_+-n_{x+}}(-i)^{n_--(n_x-n_{x+})}}{2^{n_++n_-}}\,,
\end{equation}
and for the reverse transformation
\begin{equation}
 (n_x,n_y,n_z)=\sum_{n_+=0}^{n_x+n_y}A_{n_x,n_y}^{n_+,n_-=n_x+n_y-n_+} (n_z,n_+,n_-)\,,
\end{equation}
with
\begin{equation}
 A_{n_x,n_y}^{n_+,n_-=n_x+n_y-n_+} = \frac{n_+!n_-!}{n_x! n_y!}\sum_{n_{+x}=\max[0,n_+-n_y]}^{\min[n_x,n_+]}\binom{n_x}{n_{+x}}\binom{n_y}{n_+-n_{+x}}(-i)^{n_+-n_{+x}}i^{n_y-(n_+-n_{+x})}\,.
\end{equation}

\subsection{More complex observables: QFI and entanglement entropies}

Calculating more complex observables such as quantum Fisher information and R\'enyi- and von-Neumann entanglement entropies requires to determine the eigenvalues and eigenvectors of the density matrix. It is not obvious how this can be done given the coefficient vector in the symmetrized basis. In the following, we outline how to adopt the procedure described in Ref.~\cite{Xu2013} to accomplish this. The main idea is that any permutation-symmetric state is block-diagonal if written in the Dicke basis of angular momentum states $\ket{J,M,\beta}$, where $J=N/2,N/2-1\ldots 0$ (or $1/2$), $M=-J/2\ldots J/2$, and $\beta=1\ldots n_{N,J}$, where $n_{N,J}$ is the degeneracy factor of each $J,M$ pair. By block-diagonal we mean that $\bra{J^\prime M^\prime \beta^\prime}\rho\ket{JM\beta}=\delta_{J^\prime J}\delta_{\beta^\prime \beta}\bra{J M^\prime \beta}\rho\ket{JM\beta}$. Also, blocks with different $\beta$ but same $J$ are identical, and thus one only has to calculate one representative element of each block for each $J$ and keep track of the degeneracy factor $n_{N,J}$. This means that the number of matrix-elements that have to be determined is again $\sim N^3$, in fact the number of non-redundant and non-zero matrix elements is exactly the same as in any other basis, namely $\binom{N+3}{3}$.

In Ref.~\cite{Xu2013}, Xu et al.\ derive a method to recursively construct the matrix elements in the $\ket{JM}$ basis from a symmetrized basis that is constructed with the basis operators $\ket{\su}\bra{\su}=(1+\sigma_z)/2$, $\ket{sd}\bra{sd}=(1-\sigma_z)/2$, $\ket{\su}\bra{sd}=\sigma_+$, $\ket{\sd}\bra{\su}=\sigma_-$. Thus, before we can apply the recursion, we have to transform the coefficient vector from the $(n_z,n_+,n_-)$ basis to this $(n_\su,n_\sd,n_+,n_-)$ basis. It is straight forward to show that this is accomplished by  
\begin{equation}
 c_{(n_\su,n_\sd,n_+,n_-)}=\sum_{n_z=0}^{n_\su+n_\sd} A_{n_1 n_z}^{n_\su n_\sd} c_{(n_z,n_+,n_-)} \quad \mathrm{where} \quad A_{n_1 n_z}^{n_\su n_\sd} = \frac{n_\su!n_\sd!}{n_1!n_z!}\sum_{p=\max(0,n_\su-n_z)}^{\min(n_1,n_\su)}(-1)^{n_z-n_\su+p} \binom{n_1}{p}\binom{n_z}{n_\su-p}\,,
\end{equation}
and for the reverse transformation:
\begin{equation}
 c_{(n_z,n_+,n_-)}=\sum_{n_\sd=0}^{n_1+n_z} A_{n_\su n_\sd}^{n_1 n_z} c_{(n_\su,n_\sd,n_+,n_-)} \quad \mathrm{where} \quad  A_{n_\su n_\sd}^{n_1 n_z} = \frac{n_1!n_z!}{n_\su!n_\sd!}\sum_{p=\max(0,n_z-n_\sd)}^{\min(n_z,n_\su)}(-1)^{n_z-p} \binom{n_\su}{p}\binom{n_\sd}{n_z-p}\,.
\end{equation}
Note that the coefficient vector obtained in this way corresponds to symmetrized basis states that are not normalized in the same way as those in Ref.~\cite{Xu2013}. In order to obtain the same normalization, we have to multiply each coefficient $c_{(n_\su,n_\sd,n_+,n_-)}$ by the number possible permutations $\frac{N!}{n_\su!n_\sd!n_+!n_-!}$. 

Now, finding the representation of $\rho$ in the basis $\ket{JM}\bra{JM^\prime}$, means that we want to express the coefficients $d_{JMM^\prime}=\bra{JM}\rho\ket{JM^\prime}$ in terms of the coefficients $c_{(n_\su,n_\sd,n_+,n_-)}$. 
Before we outline the recursive method, we note that the basis states of both bases are eigenstates of the superoperators $\mathcal{S}_L[\cdot]=S_z \cdot$ and $\mathcal{S}_R[\cdot]= \cdot S_z$, where $S_z=\sum_i \sigma_z^{(i)}/2$
\begin{equation}
\begin{aligned}
 2S_z (n_\su,n_\sd,n_+,n_-) & = (n_\su-n_\sd + n_+ - n_-)(n_\su,n_\sd,n_+,n_-) \\
 (n_\su,n_\sd,n_+,n_-) 2S_z & = (n_\su-n_\sd - n_+ + n_-)(n_\su,n_\sd,n_+,n_-) \\
 2S_z \ket{JM}\bra{JM^\prime} &= 2M\ket{JM}\bra{JM^\prime} \\
 \ket{JM}\bra{JM^\prime}2S_z &= 2M^\prime\ket{JM}\bra{JM^\prime}
\end{aligned}
\end{equation}
Since two eigenstates with different eigenvalues are orthogonal to each other, this means that $c_{(n_\su,n_\sd,n_+,n_-)}$ can only contribute to $d_{JMM^\prime}$ if the two constraints $M+M^\prime=n_\su-n_\sd$ and $M-M^\prime=n_+-n_-$ are fulfilled. This reduces the number of summands that contribute to $d_{JMM^\prime}$ to $\leq N$. The second constraint reflects the fact that in both bases one can uniquely assign a coherence order $m=M-M^\prime=n_+-n_-$ to each basis state, i.e., in both representation there is a natural division of the density matrix into blocks of different coherence order.

We now outline the recursive procedure for determining the basis transformation matrix.
First, we notice that $\bra{N/2\,N/2}\rho\ket{N/2\,N/2}=\bra{\su\ldots \su}\rho\ket{\su\ldots \su} = c_{(N000)}$, so we already know the first row of our basis transformation matrix. To get the next element  $\bra{N/2\,N/2}\rho\ket{N/2\,N/2-1}$ we notice that using the properties of the angular momentum eigenstates $\ket{JM}$
\begin{equation}
\begin{aligned}
 & \bra{N/2\,N/2}\rho \ket{N/2\,N/2-1} = \frac{\bra{N/2\,N/2}\rho S_-\ket{N/2\,N/2}}{\sqrt{(J+M)(J-M+1)}}  \\
  &= \sum_{n_\su,n_\sd,n_+,n_-}{c_{(n_\su,n_\sd,n_+,n_-)}} \frac{\bra{N/2\,N/2}(n_\su,n_\sd,n_+,n_-) S_-\ket{N/2\,N/2}}{\sqrt{(N/2+N/2)(N/2-N/2+1)}} \\
  &= \sum_{n_\su,n_\sd,n_+,n_-}{c_{(n_\su,n_\sd,n_+,n_-)}} \frac{\bra{N/2\,N/2}[n_\sd(n_\su,n_\sd-1,n_+,n_-+1)+n_+ (n_\su+1,n_\sd,n_+-1,n_-)]\ket{N/2\,N/2}}{\sqrt{N}} \\
  &= c_{(N-1,0,1,0)}/\sqrt{N}
\end{aligned}
\end{equation}
where $S_-=\sum_i \sigma_-^{(i)}$. We have applied $S_-$ form the right to each basis state and then made use of our knowledge of the previously calculated row of the matrix, i.e., we match the shifted indices of the coefficients to the non-zero entries of the previous row. This way, we can recursively calculate all other elements $d_{N/2\, N/2\, M}$. Then, exploiting that the density matrix is Hermitian, we get $d_{N/2\, M\, N/2}=d_{N/2\, N/2\, M}$ from which we can calculate $d_{N/2\, M\, M^\prime}$ by applying the same recursion as before starting with $d_{N/2\, M\, N/2}$.

The next step is to calculate the matrix elements for other blocks with $J<N/2$. For this we take the trace on both sides of
\begin{equation}
 \rho=\sum_{n_\su,n_\sd,n_+,n_-}c_{(n_\su,n_\sd,n_+,n_-)} (n_\su,n_\sd,n_+,n_-) = \sum_{J,M,M^\prime} d_{JMM^\prime} \ket{JM}\bra{JM^\prime} n_{N,J}
\end{equation}
where the sum on left is constrained by $N=n_\su+n_\sd+n_++n_-$. The basis states with non-zero trace are the ones with $n_+=n_-=0$ on the left and $M=M^\prime$ on the right. Using in addition that coefficients of basis states with different $\mathcal{S}_L$ (or $\mathcal{S}_R$) eigenvalues do not depend on each other, we find
\begin{equation}
\begin{aligned}
 \sum_{n_\su,n_\sd}c_{(n_\su,n_\sd,0,0)} &= \sum_{J,M} d_{JMM} n_{N,J} \\
 \Leftrightarrow \sum_{M=-N/2}^{N/2}c_{(N/2+M,N/2-M,0,0)} &= \sum_{M=-N/2}^{N/2}\sum_{J=|M|}^{N/2} d_{JMM} n_{N,J} \\
 \Leftrightarrow c_{(N/2+M,N/2-M,0,0)} &= \sum_{J=|M|}^{N/2} d_{JMM} n_{N,J} \quad \forall M \\
 \Leftrightarrow d_{|M|MM}  &= \frac{1}{n_{N,|M|}}\left(c_{(N/2+M,N/2-M,0,0)}  - \sum_{J=|M|+1}^{N/2} d_{JMM} n_{N,J} \right) \quad \forall M
\end{aligned}
\end{equation}
With this relation we can calculate $d_{JJJ}$ using that we already know $d_{J^\prime JJ}$ with $J^\prime > J$.
For example $d_{N/2-1\,N/2-1\,N/2-1}=(c_{(N-1,1,0,0)}-d_{N/2,N/2,N/2})/(N-1) = (c_{(N-1,1,0,0)}-c_{(N,0,0,0)})/(N-1)$.
Form here, we can then again recursively calculate all $d_{N/2-1\,M,M^\prime}$ as described above.
The degeneracy factor $n_{N,J}$ can be obtained from a Young tableau. A closed form expression is given by
\begin{equation}
 n_{N,J} = \frac{N! (2 J + 1)}{(N/2 + J + 1)! (N/2 - J)!}
\end{equation}

Given the density matrix of $\rho$ in the $\ket{JM}\bra{JM^\prime}$ basis, we can now diagonalize each block giving the eigenvalues $\lambda_{J,i}$ and eigenstates $\ket{\psi_{J,i}}$ and calculate the QFI \cite{braunstein1994}
\begin{equation}
 F_Q(\rho,A)=2\sum_{k,l}\frac{(\lambda_k-\lambda_l)^2}{\lambda_k+\lambda_l} |\bra{\psi_{k}}A\ket{\psi_{l}}|^2 =2\sum_J n_{N,J}\sum_{k,l}\frac{(\lambda_{J,k}-\lambda_{J,l})^2}{\lambda_{J,k}+\lambda_{J,l}} |\bra{\psi_{J,k}}A\ket{\psi_{J,l}}|^2 
\end{equation}
where in the last expression the indices $k$ and $l$ run over the size $2J+1$ of the respective block.

In order to calculate entanglement entropies, we have to calculate partial traces. This is almost trivial in the basis $(n_z,n_+,n_-)$. If we trace over $n$ out of $N$ particles, then a basis state will only contribute if all the $n$ particles are in state $1$. Taking the trace gives a factor $2^n$ for the normalization we use. Thus $c^{(N-n)}_{n_z,n_+,n_-}=2^n c^{(N)}_{n_z,n_+,n_-}$, where the particle number of the system is represented as an upper index. To calculate for example the von Neumann entropy of the reduced system, we then transform to the $\ket{JM}\bra{JM^\prime}$ basis, diagonalize and calculate $S_E=-\sum_J n_{N,J}\sum_k \lambda_{J,k} \log \lambda_{J,k}$. We proceed similarly for R\'enyi entropy and mutual information.

\end{widetext}

\end{document}